\documentclass[11pt,a4paper,reqno]{article}
\usepackage{amssymb}
\usepackage{mathrsfs}
\usepackage{graphicx}
\usepackage{flafter}
\usepackage{amsmath}
\usepackage{fancyhdr}
\usepackage{makeidx}
\usepackage{textcomp}
\usepackage{makeidx}

\newcommand{\norm}[1]{\left\Vert#1\right\Vert}
\newcommand{\abs}[1]{\left\vert#1\right\vert}
\newcommand{\set}[1]{\left\{#1\right\}}
\newcommand{\Real}{\mathbb R}
\newcommand{\eps}{\varepsilon}
\newcommand{\To}{\longrightarrow}
\newcommand{\BX}{\mathbf{B}(X)}
\newcommand{\A}{\mathcal{A}}
\begin{document}

\title{Lower Bound Approximation  to Basket Option Values for Local Volatility Jump-Diffusion Models}
\author{Guoping Xu\thanks{Risk Analytics, Citigroup Centre, Canary Wharf, 
London E14 5LB, UK.
Email: guoping.xu@citi.com}
\ and Harry Zheng\thanks{Department  of Mathematics,
Imperial College, London SW7 2AZ, UK.
Email:  h.zheng@imperial.ac.uk}
}
\date{}
\maketitle

\noindent
{\bf Abstract}. \ 
In this paper we derive  an easily computed approximation to European basket call prices
for  a local volatility jump-diffusion model. We  apply the asymptotic expansion method to find the approximate value of the lower bound of European basket call prices.
  If the
local volatility function is time independent then there is a closed-form
expression for the approximation. Numerical tests  show that the suggested approximation is  fast and accurate 
 in comparison with the Monte Carlo and other approximation methods  in the literature.

\bigskip\noindent{\bf Keywords}. \
Basket options valuation, local volatility jump-diffusion model, 
lower bound approximation, second order asymptotic expansion. 



\section{Introduction}

It is in general difficult to value basket options due to lack of analytic characterization of the distribution of the underlying basket asset price process. Research is mainly focused on developing fast  and accurate approximation methods and finding tight  lower and upper bounds for basket option values.   In the Black-Scholes setting Curran (1994) and Rogers and Shi (1995) derive a lower bound for Asian options by the conditioning random variable and Jensen's inequality.
 Deelstra et al.~(2004) obtain the bounds for basket options with the comonotonicity approach.  In 
affine L\'evy models one can derive lower bounds numerically for arithmetic
Asian options based on  the characteristic function and the method developed in Duffie et al.~(2000), see Albrecher et al.~(2008). Deelstra et al.~(2010) provide a good overview of the recent development in finding and computing the bounds.     

It is known that  Rogers and Shi's lower bound is generally  tight and is one of the
most accurate approximations to basket call option values. 
The lower bound can be calculated exactly in the Black-Scholes framework.
Xu and Zheng (2009) show that the lower bound can also be calculated exactly
in a special jump-diffusion model with constant volatility and two types of Poisson jumps
(systematic and idiosyncratic jumps).  
The usefulness of  Rogers and Shi's lower bound depends
  crucially on one's ability of finding some highly correlated random variables to the basket value and computing the conditional expectation exactly.
It is difficult to extend Rogers and Shi's lower bound to more general
models such as local volatility models due to lack of explicitly known distributions
  for  models with non-constant volatilities, see Albrecher et al.~(2008).
   To the best of our knowledge,  Rogers and Shi's lower bound for models with local volatilities
   has not been discussed in the literature.

In this paper we aim to find  a good approximation to Rogers and Shi's lower
bound for  a local volatility jump-diffusion model and  use it 
to approximate  European basket call option values. 
 We first apply the second order asymptotic expansion (see  Benhamou et al.~(2009)) to approximate the basket asset value, then choose a normal variable and a
Poisson variable as  conditioning
variables which are highly correlated to the basket asset value, and finally apply   the conditional
expectation results of multiple Wiener-It\^o integrals (see  Kunitomo and Takahashi (2001))  to approximate Rogers and Shi's lower bound.   The main contribution of the paper is the
derivation of an approximation to Rogers and Shi's lower bound 
for  local volatility jump-diffusion models. We
suggest an easily implemented algorithm   to compute the lower bound approximation. If the
local volatility function is time independent then there is a closed-form
expression for the approximation. Numerical tests show that the lower bound approximation is  fast and accurate 
in most cases in comparison with the Monte Carlo method, the partial-exact approximation  (Xu and Zheng (2009)) and 
the asymptotic expansion approximation (Xu and Zheng (2010)).

The paper is organized as follows. Section 2 formulates the basket local volatility jump-diffusion
 model and explains some known methods in pricing European basket call options. Section 3 
applies the second order asymptotic expansion to derive an  easily computed approximation to Roger and Shi's
lower bound. Section 4 
compares the numerical performance of the lower bound approximation with
other methods in the literature. Section 5 concludes.

\section{Model}
Assume  $(\Omega,P,{\cal F}, {\cal F}_t)$ is a filtered risk-neutral
 probability space
and ${\cal F}_t$ is the augmented natural filtration generated by 
correlated Brownian motions $W_1,\ldots, W_n$ with correlation matrix
$(\rho_{ij})$ 
and a Poisson process $N$ with intensity $\lambda$.
Assume  Brownian motions $W_1,\ldots,W_n$ 
and Poisson process $N$ are independent to each other.
Assume the riskfree interest rate is zero for the sake of clarity (otherwise one may simply consider the discounted asset prices and the discounted strike price). 
Assume the portfolio is composed of $n$ assets 
with prices $S_1,\ldots, S_n$ that are martingales 
satisfying the following stochastic differential equation
\begin{align}{\label{eq100222.1}}{d S_i(t)}=\sigma_i(t,S_i(t))d
W_i(t)+h_iS_i(t-)dM(t),
\end{align}
where  $h_i>-1$ are constant jump sizes (percentage changes of the values at the jump time of $N$), $\sigma_i(t,S_i)$ are the local volatility functions,  
and $M(t)=N(t)-\lambda t$ is the compensated Poisson process.  
The basket value at time $T$ is given by
\begin{align*}S(T)=\sum_{i=1}^{n}w_iS_i(T),
\end{align*}
where $w_i$ are positive constant weights. The  European basket call option price at time 0 is given by
\begin{align*}C(T,K)=\mathbf{E}[(S(T)-K)^+].
\end{align*}

Almost all works in the literature on Asian or basket options valuation assume
the underlying asset prices  follow lognormal processes, which corresponds
to $\sigma_i(t,S_i)=\sigma_iS_i$ 
with $\sigma_i$ being positive constant and $h_i=0$ for all $i$
in (\ref{eq100222.1}).  Since Asian options are similar to basket options,  we do not differentiate
these two types of options, even though some techniques are originally 
 developed for Asian options. We now  
review some well-known approaches in approximation and error bound estimation
for the Black-Scholes case.  

Levy (1992)
approximates the basket value $S(T)$ 
with a lognormal variable which matches the  
 first two moments of $S(T)$ and derives an approximate closed-form
 pricing formula for $C(T,K)$.  The result is good when maturity $T$ and volatilities $\sigma_i$
 are relatively small. The performance deteriorates as $T$ or $\sigma_i$
 increases.   Curran (1994) introduces the idea of conditioning 
 variables. Assume $\Lambda$ 
is a random variable which has strong correlation
with $S(T)$ and satisfies that $S(T)\geq K$
whenever $\Lambda\geq d_\Lambda$ for some constant $d_\Lambda$. 
The basket 
option price can be decomposed as 
\begin{equation}\label{e2}
\mathbf{E}[(S(T)-K)^{+}]
=\mathbf{E}[(S(T)-K)1_{[\Lambda\geq
d_\Lambda]}]+\mathbf{E}[(S(T)-K)^{+}1_{[\Lambda<d_\Lambda]}].
\end{equation}
Curran (1994) chooses the geometric average of the asset prices as the conditioning variable $\Lambda$ (a lognormal variable)
and finds the closed-form expression for the the first term in (\ref{e2})
and uses the lognormal variable and  the 
conditional moment matching 
 to find the approximate value of the second term in (\ref{e2}).
Rogers and Shi (1995) 
use the conditioning variable $\Lambda$ to
show that  the   lower and upper bounds of the basket option price  $C(T,K)$
are given by
$$ LB=\mathbf{E}\left[(\mathbf{E}[S(T)|\Lambda]-K)^{+}\right]$$
and
$$UB=LB
 + {1\over 2}\mathbf{E}\left[\mathrm{var} 
(S(T)|\Lambda)1_{[\Lambda<d_\Lambda]}\right]^{\frac{1}{2}}
\mathbf{E}[1_{[\Lambda<d_\Lambda]}]^{\frac{1}{2}}.
$$
These bounds may be computed analytically for specific conditioning variables $\Lambda$. Numerical tests show that Rogers and Shi's lower bound is normally very 
close to the true (Monte Carlo) value.   

Xu and Zheng (2009) extend the Black-Scholes model of 
Curran (1994) to the jump-diffusion model (\ref{eq100222.1}) with 
local volatility functions
$\sigma_i(t,S_i)=\sigma_i
S_i$ for all $i$. (The jump part of the model in Xu and Zheng (2009) is more general than (\ref{eq100222.1}) with both common  and individual jumps.) Asset prices $S_i(T)$ at time $T$ 
have closed form expressions 
\begin{equation} \label{SiT}
 S_i(T) = S_i(0)\exp\left(-{1\over 2}\sigma_i^2T - h_i\lambda T
+\sigma_i W_i(T) + N(T)\ln(1+h_i)\right),
\end{equation}
where $S_i(0)$ are asset prices at time 0, for $i=1,\ldots,n$. We can write the basket price
$S(T)$ as
$$S(T)= \sum_{i=1}^n a_i \exp\left(\sigma_i W_i(T) + N(T)\ln(1+h_i)\right),$$
where $a_i=w_i S_i(0) 
\exp((-\frac{1}{2}\sigma_i^2-h_i\lambda)T)$, $i=1,\ldots,n$.
Since $\exp(x)\geq 1+x$ for all $x$ and $a_i\geq 0$ (weights $w_i$ nonnegative) the basket price $S(T)$  satisfies
$$S(T)\geq \sum_{i=1}^n a_i \left(1+\sigma_i W_i(T) + N(T)\ln(1+h_i)\right)
=c+ mN(T) + \sigma W,
$$
where 
$c=\sum_{i=1}^n a_i$, $m=\sum_{i=1}^n a_i \ln(1+h_i)$, $\sigma^2=\sum_{i,j=1}^n 
a_i a_j \rho_{ij} \sigma_i  \sigma_j T$, $N(T)$ is a Poisson variable with parameter $\lambda T$, and 
$ W={1\over \sigma}\sum_{i=1}^n a_i\sigma_i W_i(T)$
is a standard normal variable  and is independent of $N(T)$.
Note that the above inequality holds only when all weights $w_i$ are nonnegative, which excludes spread options or basket options with negative weights.

If we choose a conditioning variable by $\Lambda=(N(T),W)$
then   $S(T)\geq K$ whenever $c+ mN(T) + \sigma W\geq K$.  it is easy to find 
\begin{equation} \label{EST}
E[S(T)|\Lambda=(k, y)]= 
\sum_{i=1}^n a_i \exp\left({1\over 2}(\sigma_i^2T - R_i^2) + R_iy+k\ln(1+h_i)\right),
\end{equation}
where $R_i={1\over \sigma} \sum_{j=1}^n a_j\rho_{ij} \sigma_i\sigma_jT$.
The lower bound can  be  computed from the expression
$$ LB 
=\sum_{k=0}^\infty p_k \int_{-\infty}^\infty
(E[S(T)|\Lambda=( k, y)]-K)^+ d\Phi(y),  $$
where $p_k=P(N(T)=k)=\exp(-\lambda T)(\lambda T)^k/k!$ and  $\Phi$ is the cumulative distribution function of a standard
normal variable.  Xu and Zheng (2009) suggest an approximation, called the 
partial exact approximation (PEA), 
 to the basket call option value  $C(T,K)$, given by 
\begin{eqnarray*}C^A(T,K)&:=& \mathbf{E}[(S(T)-K)^{+}1_{[c+ mN(T) + \sigma W\geq K]}]\\
&&{}+\sum_{i=1}^3 q_i\mathbf{E}[(\mathbf{E}[S(T)|\Lambda]+\alpha_i-K)^{+}1_{[c+ mN(T) + \sigma W < K]}],
\end{eqnarray*}
where $q_1=1/6$, $q_2=2/3$, $q_3=1/6$, and $\alpha_1=-\sqrt{3}\epsilon_0$,
$\alpha_2=0$, $\alpha_3=\sqrt{3}\epsilon_0$, $\epsilon_0$ is some 
constant depending on the conditioning variable $\Lambda$ and the conditional
second moment of $S(T)$ given $\Lambda$. The lower bound plays a dominant role in the approximation
with a weight $2/3$, the other two parts with a weight $1/6$ each are
the second moment  adjustment to the lower bound. 
Xu and Zheng (2009) show that  $LB\leq C^A(T,K)\leq UB$, which provides the error bounds for the approximate basket option value. 
The PEA method is fast and accurate in comparison with some other
well known  numerical schemes for basket options, see Xu and Zheng (2009) for details.
A key condition for the PEA method to work is that one must know the
closed form expressions of asset prices $S_i(T)$ as in (\ref{SiT})
for $i=1,\ldots,n$.
This is impossible for general local volatility functions.

The PEA method can be easily modified to  include the possibility of defaults of some underlying names in the portfolio. For example, assume that company $n$ defaults whenever  there is a jump event, then we may set $h_n=-1$, which implies $S_n(T)=0$ if $N(T)=k\geq 1$.
The formula for conditional expectation (\ref{EST}) is still valid but we need to differentiate the cases of no jump ($k=0$) and at least one jump ($k\geq 1$). In the former  we can use the same $\sigma^2$ and $c$ as above with the convention $\exp(0\ln 0)=1$. In the latter, due to $S_n(T)=0$  we need to remove  terms with index $n$ in all expressions, e.g.,   $m=\sum_{i=1}^{n-1} a_i \ln(1+h_i)$ and $\sigma^2=\sum_{i,j=1}^{n-1} 
a_i a_j \rho_{ij} \sigma_i  \sigma_j T$.

Xu and Zheng (2010) discuss the basket options pricing
for  asset model (\ref{eq100222.1}) with 
the same jump sizes $h_i=h$ for $i=1,\ldots,n$.  
Since the PEA method cannot be applied for general local volatility functions
$\sigma_i(t,S_i)$, Xu and Zheng (2010) first reduce the dimensionality of the problem by directly working on the portfolio asset value $S(t)$ which satisfies the following
 stochastic volatility jump diffusion model 
\begin{eqnarray*}{\label{eq090728.2}}d S(t)=V(t)
dW(t)+hS(t-)dM(t)
\end{eqnarray*}
with the initial price $S(0)=\sum_{i=1}^{n}w_i S_i(0)$,
where  $W$ is a standard Brownian motion, independent of
$M$, and 
\begin{eqnarray*}V(t)^2:=\sum_{i,j=1}^{n} 
w_iw_j\sigma_i(t,S_i(t))\sigma_j(t,S_j(t))\rho_{ij}.
\end{eqnarray*}
The European basket call option price $C(T,K)$ at time 0 satisfies 
a  partial integral differential equation (PIDE) 
\begin{equation*} 
C_T(T,K)=\lambda h  KC_K(T,K)
+\frac{1}{2}\sigma(T,K)^2C_{KK}(T,K) 
 +\lambda (h+1)
\left(C(T, {K\over h+1})
 - C(T,K)\right)
\label{pide}
\end{equation*}
with the initial condition $C(0,K)=(S(0)-K)^{+}$ and the local 
variance function 
\begin{eqnarray*}\sigma(T,K)^2
&=& \mathbf{E}[V(T)^2|S(T)=K].
\end{eqnarray*}
The main difficulty
is  to compute the conditional expectation as asset prices
$S_i(T)$ have no closed-form expressions. 
Following the first order asymptotic expansion methods of 
Kunitomo and Takahashi (2001)
and Benhamou et al.~(2009), Xu and Zheng (2010) show that the unknown local variance function $\sigma(T,K)^2$
can be approximated by
\begin{equation} \label{approx_sigma}
\sigma(T,K)^2\approx a(T) + b(T)(K- S(0)),
\end{equation}
where 
$a(T)=\sum_{i,j=1}^{n} w_iw_j\rho_{ij}p_ip_j$,
$b(T)= \sum_{i,j=1}^{n} 
\frac{1}{\sigma_c^2}w_iw_j\rho_{ij}p_ip_j
\left(\frac{q_i}{p_i}C_{i}+\frac{q_j}{p_j}C_{j}\right)$,
$p_i=\sigma_i(T,S_i(0))$, 
$q_i={\partial\over \partial S_i(0)}\sigma_i(T,S_i(0))$,
 $C_{i}
=\sum_{j=1}^{n}w_j \left[\rho_{ij}(\int_0^T\sigma_i(t,S_{i}(0))\sigma_j(t,S_{j}(0))dt)
\right]$, and
$\sigma_c^2 = \sum_{i=1}^{n}w_i C_j$.
One can find the approximate basket option price by solving the PIDE 
with the implicit-explicit finite difference method. This approximation, called the  asymptotic expansion approximation (AEA), is less accurate than the PEA  when the local volatility function is the Black-Scholes type, but is capable of dealing with  general local volatility functions.
The main limitations of the AEA method are that the approximate variance function 
in (\ref{approx_sigma}) is not always positive, which can cause numerical errors in solving the PIDE, and that it requires  the same jump size for all assets, which is unrealistic for a general asset portfolio.


\section{Lower Bound Approximation}
The asymptotic expansion for basket options
pricing in a general diffusion model is discussed in Kunitomo and Takahashi (2001) in which the valuation of
conditional expectations is  a necessary step to obtain the   characteristic
function of the basket value. In this paper we   use the same method to expand the parameterized asset
price processes  to the second order, see  Benhamou et al.~(2009), 
and apply the conditional expectation results of multiple Wiener-It\^o
integrals directly
to approximate  Rogers and Shi's lower bound for the basket call option values in a jump-diffusion model
(\ref{eq100222.1}). For $\epsilon\in[0,1]$  define
\begin{align}{\label{eq100223.1}}
{d S_i^\epsilon(t)}=\epsilon\sigma_i(t,S_i^\epsilon(t)) d W_i(t)+
\epsilon h_iS_i^\epsilon(t-)dM(t)
\end{align}
with initial condition $S_i^\epsilon(0)=S_i(0)$. Note that 
$S_i^1(T)=S_i(T)$.
Define 
$$ S_{i}^{(k)}(t):=\frac{\partial^k S_i^\epsilon(t)}{\partial
 \epsilon^k}|_{\epsilon=0}
 \quad\mbox{and}\quad 
\sigma_i^{(k)}(t):=\frac{\partial^k \sigma_i(t,S_i^\epsilon(t))}{\partial
 (S_i^\epsilon)^k}|_{\epsilon=0}$$
for $k=0,1,\ldots$. It is obvious that
$S_{i}^{(0)}(t)=S_i^0(t)=S_i(0)$ and $\sigma_i^{(0)}(t)=\sigma_i(t,S_i(0))$ for
for all $t\geq 0$ and $i=1,\ldots,n$. In particular, 
if local volatility functions $\sigma_i$ are time independent then functions $\sigma_i^{(k)}$ are all constants for $i=1,\ldots,n$, which may simplify considerably some computations. An important case is the constant elasticity of variance (CEV) model, where $\sigma_i(t,S_i)=\alpha_i S_i^{\beta_i}$ for $i=1,\ldots,n$.
To simplify the notations in subsequent discussions we denote by
\begin{equation*}
\tilde\sigma_i^{(k)}(t) = \sum_{j=1}^n w_j \sigma_i^{(k)}(t)
\sigma_j^{(0)}(t) \rho_{ij},\quad k=0,1.
\end{equation*}

The second order asymptotic expansion
 around $\epsilon=0$ for $S_i^\epsilon(t)$ is 
\begin{equation*}{\label{eq100216.5}}
S_{i}^{\epsilon}(T) \approx
S_{i}^{(0)}(T)+S_{i}^{(1)}(T)\epsilon
+\frac{1}{2}S_{i}^{(2)}(T){\epsilon}^2.
\end{equation*}
Expand \eqref{eq100223.1} to the  second order, we have
\begin{align*}
&d S_{i}^{(1)}(t)=\sigma_i^{(0)}(t) d W_i(t)+h_iS_i(0)dM(t),\\
&d S_{i}^{(2)}(t)=2\sigma_i^{(1)}(t) S_{i}^{(1)}(t)d W_i(t)+2 h_iS_{i}^{(1)}(t-)dM(t),
\end{align*}
with the initial conditions  $S_{i}^{(1)}(0)=S_{i}^{(2)}(0)=0$. Therefore, 
\begin{eqnarray}
S_{i}^{(1)}(t)&=&-\lambda h_iS_i(0)t+\int_{0}^{t}\sigma_i^{(0)}(s) d W_i(s)+h_iS_i(0)N(t) \label{Si1t}\\
S_{i}^{(2)}(t)&=&-2\lambda h_i \int_{0}^{t}S_{i}^{(1)}(s)ds
+2\int_{0}^{t}\sigma_i^{(1)}(s) S_{i}^{(1)}(s)d W_i(s)+2 h_i\int_0^t S_{i}^{(1)}(s-) dN(s) \label{Si2t}
\end{eqnarray}
for $0\leq t\leq T$. Letting $\epsilon=1$ we may approximate
the basket value $S(T)$ by
\begin{equation} {\label{eq100216.9}}
S(T)\approx S^A(T):=
S(0)+S^{(1)}(T)+\frac{S^{(2)}(T)}{2},
\end{equation} 
where $S^{(j)}(T):=\sum_{i=1}^{n}w_iS_{i}^{(j)}(T)$ for $j=1, 2$. 

 Since there are no closed-form expressions for $S_i(T)$ and $S(T)$,
it is difficult to compute the lower bound. If we approximate $S(T)$
by $S^A(T)$, defined in ({\ref{eq100216.9}}), we may be able to compute
the conditional expectation 
$\mathbf{E}[S^A(T)|\Lambda]$ for some conditioning variable $\Lambda$
and then to approximate the lower bound. We therefore propose that 
\begin{align} \label{LBA} 
 LB\approx 
 LBA :=\mathbf{E}[(\mathbf{E}[S^A(T)|\Lambda]-K)^+]. 
\end{align} 
This approximation is called the lower bound approximation (LBA) of the basket call option price. Note that the LBA is only an approximation to Rogers and Shi's lower bound and is therefore possible to have values greater than the basket call option values. 

The next step is to choose the conditioning variable for the approximation. 
 In the Black-Scholes framework and for European arithmetic average options, it is without exception to choose the geometric average of asset prices as the conditioning variable. This choice is natural as the geometric average of lognormal variables is again a lognormal variable and provides good information on the arithmetic average.  It is not clear what one
should choose for general jump-diffusion models as there are no closed form
solutions for underlying asset prices. With the insight from 
 Xu and Zheng (2009) 
we choose $\Lambda=(N(T),\Delta(T))$  for the model, where
$\Delta(T)=\sum_{j=1}^n w_j\int_{0}^{T}\sigma^{(0)}_j(t) d W_j(t)$
is a normal variable  with mean 0 and variance
$v^2=\sum_{i=1}^n w_i \int_0^T \tilde\sigma^{(0)}_i(t) dt$, 
and $N(T)$ is a Poisson variable  with parameter $\lambda T$.
From (\ref{LBA}) and (\ref{eq100216.9}) we have
\begin{align}
LBA
=&\sum_{k=0}^{\infty}p_k\int_{-\infty}^{\infty}\bigg[\bigg(S(0)
+\mathbf{E}[S^{(1)}(T)|\Lambda=(k,vx)]
+\mathbf{E}[\frac{S^{(2)}(T)}{2}|\Lambda=(k,vx)]
-K\bigg)^+\bigg]d\Phi(x), {\label{eq100228.21}}
\end{align}  
where $p_k=\exp(-\lambda T) (\lambda T)^k/k!$. Since $S^{(j)}(T)=\sum_{i=1}^n w_i S_{i}^{(j)}(T)$, we only need to find
$\mathbf{E}[S_{i}^{(j)}(T)|\Lambda=(k,vx)]$ for $j=1,2$ and $i=1,\ldots,n$.
We first calculate $\mathbf{E}[S_{i}^{(1)}(T)|\Lambda=(k,vx)]$.
\begin{align*}
&\mathbf{E}[S_{i}^{(1)}(T)|\Lambda=(k,vx)]\\
=&-\lambda h_iS_i(0)T+\mathbf{E}[h_iS_i(0)N(T)|N(T)=k]
+\mathbf{E}\bigg[\int_{0}^{T}\sigma_i^{(0)}(t) d W_i(t)|\Delta(T)=vx\bigg]\\
=&-\lambda h_iS_i(0)T+h_iS_i(0)k+
{1\over v}\left(\int_0^T \tilde\sigma_i^{(0)}(t)dt\right) x.
\end{align*}  

The valuation of $\mathbf{E}[\frac{S^{(2)}(T)}{2}|\Lambda=(k,vx)]$ 
is more involved.
Since $S_i^{(2)}(T)$ is the sum of  three terms in (\ref{Si2t}) we may 
find  the conditional expectation of each term by substituting $S_{i}^{(1)}(t)$ in (\ref{Si1t})
into the integrands and then computing three conditional expectations. 
We now derive them one by one. The first term can be written as
$$ (-\lambda h_i)\mathbf{E}\bigg[ \int_{0}^{T}S_{i}^{(1)}(t)dt|\Lambda=(k,vx)\bigg]
=A_1+A_2+A_3,$$
where
\begin{eqnarray*}
A_1&=& (-\lambda h_i)
\mathbf{E}\bigg[\int_{0}^{T}(-\lambda h_iS_i(0)t)dt|\Lambda=(k,vx)\bigg]\\
&=& \frac{1}{2} S_i(0)\lambda^2 h_i^2T^2\\
A_2 &=&(-\lambda h_i)
\mathbf{E}\bigg[\int_{0}^{T}(\int_{0}^{t}\sigma_i^{(0)}(s) d W_i(s)
)dt|\Lambda=(k,vx)\bigg]\\
&=&(-\lambda h_i)\mathbf{E}\bigg[\int_{0}^{T}(T-t)\sigma_i^{(0)}(t) d W_i(t)|\Delta(T)=vx\bigg]\\
&=&(-\lambda h_i){1\over v}\left(\int_0^T (T-t)\tilde\sigma_i^{(0)}(t) dt\right) x\\
A_3 &=& (-\lambda h_i)
\mathbf{E}\bigg[\int_{0}^{T}(h_iS_i(0)N(t))dt|\Lambda=(k,vx)\bigg]\\
&=&(-\lambda S_i(0)h_i^2)\int_{0}^{T}\mathbf{E}[N(t)|N(T)=k]dt\\
&=&(-\lambda S_i(0)h_i^2)\frac{(kT)}{2}.
\end{eqnarray*}
Here we have used the fact that 
$(N(t)|N(T)=k)$ is a binomial variable with k independent 0-1 trials and
probability $\frac{\lambda t}{\lambda T}=\frac{t}{T}$
of taking 1, which implies 
$\mathbf{E}[N(t)|N(T)=k]=\frac{kt}{T}$.
The second term can be written as
$$ \mathbf{E}\bigg[\int_{0}^{T}\sigma_i^{(1)}(t) S_{i}^{(1)}(t)d W_i(t)|\Lambda=(k,vx)\bigg]
= B_1+B_2+B_3,$$
where
\begin{eqnarray*}
B_1 &=& \mathbf{E}\bigg[\int_{0}^{T}\sigma_i^{(1)}(t)\left(-\lambda h_iS_i(0)t\right)d W_i(t)|\Lambda=(k,vx)\bigg]\\
&=& (-\lambda h_i)S_i(0)\bigg(\int_{0}^{T}t\tilde\sigma_{i}^{(1)}(t) d t\bigg)\frac{1}{v}x\\
B_2 &=& \mathbf{E}\bigg[\int_{0}^{T}\sigma_i^{(1)}(t)
\left(\int_{0}^{t}\sigma_i^{(0)}(s) d W_i(s)\right)d W_i(t)|\Lambda=(k,vx)\bigg]\\
&=& \frac{1}{v^2}\left(\int_{0}^{T}\left(\int_{0}^{t}\tilde\sigma_i^{(0)}(s) ds\right)\tilde\sigma_i^{(1)}(t) dt\right)(x^2-1)\\
B_3 &=& \mathbf{E}\bigg[\int_{0}^{T}\sigma_i^{(1)}(t)
\left(h_iS_i(0)N(t)\right)d W_i(t)|\Lambda=(k,vx)\bigg]\\
&=&h_iS_i(0)\mathbf{E}\bigg[\int_{0}^{T}\sigma_i^{(1)}(t)
\frac{kt}{T}d W_i(t)|\Delta(T)=vx\bigg]\\
&=&h_iS_i(0)\frac{k}{T}\bigg(\int_{0}^{T}t\tilde\sigma_{i}^{(1)}(t) 
 d t\bigg)\frac{1}{v}x.
\end{eqnarray*}
The computation of $B_2$ is discussed  in 
 Kunitomo and Takahashi (2001,  Lemma A.1).
The third term can be written as
$$h_i\mathbf{E}\bigg[\int_0^T S_{i}^{(1)}(t-) dN(t)|\Lambda=(k,vx)\bigg]
=C_1+C_2+C_3,$$
where
\begin{eqnarray*}
C_1&=&h_i\mathbf{E}\bigg[\int_0^T \bigg(-\lambda h_iS_i(0)t
\bigg) dN(t)\bigg|\Lambda=(k,vx)\bigg]\\
&=& -\frac{1}{2}S_i(0)\lambda h_i^2Tk\\
C_2 &=& h_i\mathbf{E}\bigg[\int_0^T \bigg(
\int_{0}^{t}\sigma_i^{(0)}(s) d W_i(s)\bigg) dN(t)\bigg|\Lambda=(k,vx)\bigg]\\
&=& h_i\mathbf{E}[\int_{0}^{T}\bigg(\int_{0}^{t}\sigma_i^{(0)}(s) d W_i(s)\bigg)\frac{k}{T}dt|\Delta(T)=vx]\\
&=& h_i\frac{k}{T}\bigg(\int_{0}^{T}(T-t)\tilde\sigma_{i}^{(0)}(t)  d t\bigg)\frac{1}{v}x\\
C_3 &=& h_i\mathbf{E}\bigg[\int_0^T \bigg(
h_iS_i(0)N(t-)\bigg) dN(t)\bigg|\Lambda=(k,vx)\bigg]\\
&=& h_i^2S_i(0)\mathbf{E}\bigg[\sum_{l=0}^{N(T)-1}l|N(T)=k\bigg]\\
&=&h_i^2S_i(0)\frac{k^2-k}{2}.
\end{eqnarray*}

Substituting the first and second order conditional expectations into 
(\ref{eq100228.21}) we get the lower bound approximation as
\begin{equation} \label{lb}
 LBA = \sum_{k=0}^{\infty} p_k
\int_{-\infty}^{\infty}\bigg(cx^2+a_1(k)x + a_0(k)\bigg)^+d\Phi(x),
\end{equation}
where 
\begin{align}
c&=\frac{1}{v^2} \sum_{i=1}^n w_i \left(\int_{0}^{T}\left(\int_{0}^{t}\tilde\sigma_i^{(0)}(s) ds\right)\tilde\sigma_i^{(1)}(t) dt\right)\nonumber \\
a_0(k)&=S(0)+(K-\lambda T)\sum_{i=1}^n w_ih_i S_i(0)
+\frac{1}{2}\sum_{i=1}^{n}w_i S_i(0)h_i^2\big((k-\lambda T)^2-k\big)-c-K\nonumber\\
a_1(k)&=v+\frac{1}{v}\sum_{i=1}^{n}w_i h_i(\frac{k}{T}-\lambda)
\bigg(\int_{0}^{T}\big((T-t)\tilde\sigma_{i}^{(0)}(t)+S_i(0)
t\tilde\sigma_i^{(1)}(t)\big)  d t\bigg).\nonumber
\end{align}
Note that in computing the  lower bound approximation (\ref{lb}) there is no need to do numerical integration. This is because for every fixed $k$ the integrand is the positive part of a quadratic function which contains only three cases: no root, one root, and two roots.  Therefore the computation of the  lower bound  approximation $LBA$ is easy and fast. This is one key advantage over the other methods such as simulation or finite difference for the PIDE.

\section{Numerical Results}
In this section we conduct some numerical tests for European basket calls
with underlying asset price processes satisfying
(\ref{eq100222.1}) to test
the performance of the lower bound approximation.  The Monte Carlo simulation
provides the benchmark results. The control variate technique is adopted
to reduce the standard deviations. The control variate used is the first order asymptotic expansion of the basket price $S_(T)$, which results in a closed form pricing formula, see Xu and Zheng (2010) for details. 
The following data are used in all numerical tests:
the number of assets in the basket $n=4$,
the portfolio weights  of each asset $w_i=0.25$ for $i=1,\ldots,n$, 
the correlation coefficients of Brownian motions 
$\rho_{ij}=0.3$ for $i,j=1,\ldots,n$, the initial asset
prices $S_i(0)=100$ for $i=1,\ldots,n$,
the exercise price $K=100$, and the jump intensity $\lambda=0.3$. The number of simulations for each test case is 100,00 and the Poisson summation is approximated by truncating the infinite series  after the first 10 terms.

Tables \ref{table3.4}   displays the European basket call option values and their implied volatilities with four different methods: the Monte Carlo (MC), the
partial exact approximation (PEA), the asymptotic expansion (AEA)
and the lower bound approximation (LBA). The local volatility function is  $\sigma(t,S)=0.2S $. We perform
numerical tests for three constant jump sizes  
$h_i=e^\eta-1$ with $\eta=-0.25$, $-0.125$ and $-0.0625$. 
 It is clear that all three approximation methods perform well with  relative errors less than 1\%. We have done other tests with the same data except the volatility function being changed to $\sigma(t,S)=0.5S$. The overall relative error is 0.6\% for the PEA, 4\% for the AEA  and 1.7\% for the LBA. 
Tables \ref{table3.4}  shows that the performance of the PEA  is the best 
while the LBA performs better than the AEA. 
The results suggest  that the PEA is the best approximation method for the European basket call  option valuation when  local volatility functions are of the Black-Scholes type. On the other hand, the AEA and LBA  are much more flexible and can handle general local volatility functions. The LBA has the additional advantage of having
closed-form solutions when the local volatility functions are time independent and can deal with  different jump sizes to common jumps.
The AEA  is
slow in solving the PIDE with the finite difference method and requires the same jump size to common jumps for all assets. It is also interesting to note that the implied volatilities are essentially same for $T=1$ and  $T=3$ with all other parameters being the same. It is not yet clear to us the reason of  insensitivity of the implied volatility with respect to the time to maturity.

Tables \ref{table2}  displays the European basket call option values and their implied volatilities with the Monte Carlo (MC) and the lower bound approximation (LBA).
This test is to compare the results with different moneyness ($K/S$) from deep in the money 0.7 to deep out of the money 1.3. The other parameters are also being set near the extreme case to see the numerical approximation performance, including the jump intensity $\lambda=4$ (frequent jumps), correlation coefficient of Brownian motions $\rho=0.9$ (highly correlated),  and local volatility function $\sigma(t,S)=0.5S$ (high volatility). Jump sizes of four assets are also different to reflect the heterogeneous setting with $h_1=0$, $h_2=0.1$, $h_2=0.3$, $h_3=-0.5$, which implies that jump has no impact to asset 1, small positive impact to assets 2 and 3, and big negative impact to asset 4.  We compute the prices and their implied volatilities. The average relative errors are about 4 percent, which is heavily influenced by errors in computing out-of-money call option prices with long maturity ($T=2$).
This is perhaps not surprising as the asymptotic expansion method works  well only when all coefficients are very small whileas the parameters in this test  are chosen to be large.

Table \ref{table3.1}  displays the numerical results 
with the MC, AEA and LBA and with different maturities
($T=1,3$) and  local volatility functions
($\sigma(S)=\alpha S^{\beta}$ with $\alpha=0.2,0.5$ and $\beta=1,0.8,0.5$). 
  The jump size is $h_i=e^{-0.25}-1=-0.2212$ for all assets. The last row displays 
the average errors of the AEA and LBA. 
Table \ref{table3.1} shows that
the performance of the LBA  is excellent with the average relative error
0.4\%. The overall performance of the LBA  is better than
that of the AEA,  especially  when
the local volatility function is $\sigma(S)=0.5S$.   Matlab is used for computation.
The LBA only takes a few seconds for each case,  much faster than the AEA and MC. 
We have done  other tests with the same data as in Table \ref{table3.1}
except different intensity rates ($\lambda=0$  and 1). The overall performance is essentially the same as that of Table \ref{table3.1}. We can say with reasonable confidence that the LBA  suggested in this paper works well.

 Tables \ref{table3.3}   displays the numerical results with the MC and
LBA and  with different maturities $T$
and  local volatility functions  $\sigma(t,S)=\alpha S^\beta$. The jump sizes for four assets
are $h_1=0$, $h_2=0.3$, $h_3=-0.3$, and $h_4=0$, which implies that the jump event has no impact to asset prices 1 and 4, positive impact to asset price 2 and negative impact to asset price 3. Since The PEA  requires the Black-Scholes setting while the AEA  requires the same jump size for all assets, neither method can be used in this test.  
It is clear that the LBA   performs well in comparison with  the MC with the overall relative error less than 1\%.

\section{Conclusions}
In this paper we  suggest to approximate the basket call option value of a local volatility jump-diffusion model with  Rogers and Shi's lower bound which is approximated with the second order asymptotic expansion method.  The lower bound approximation (LBA) is easily computed and, If the
local volatility function is time independent (e.g., the CEV model), has a closed-form
expression. We  compare the numerical performance of the LBA with other   methods using different parameters and  show that the LBA  is fast and accurate in most
cases.  

\bigskip
\noindent{\bf Acknowledgment.}
The authors are grateful to the anonymous referees for their constructive comments and suggestions which have helped to improve the paper.

\begin{table}
\begin{center}
{\small
\renewcommand{\arraystretch}{1.2}
\begin{tabular}{|c|c|c|rrrr|}
\hline
$\lambda$& $\eta$& $T$ & MC (stdev)  & PEA (IV\%) & AEA (IV\%) & LBA (IV\%)\\ \hline
0.3&$-0.25$&1&7.35 (0.01)&7.35 (18.4) &7.35 (18.4)&7.37 (18.5) \\
\cline{3-7}
&&3&12.93 (0.01) &12.92 (18.8)&12.85 (18.7)&12.86 (18.7)\\
\cline{2-7}
&$-0.125$&1&6.08 (0.01) &6.08 (15.3)&6.07 (15.2)&6.09 (15.3)\\
\cline{3-7}
&& 3&10.57 (0.01) &10.56 (15.3)& 10.49 (15.2)& 10.57 (15.3)\\
\cline{2-7}
& $-0.0625$ & 1&5.66 (0.01)&5.66 (14.2)&5.65 (14.2)&5.67 (14.2)\\
\cline{3-7}
&& 3&9.83 (0.01)&9.82 (14.2)& 9.74 (14.1)&9.86 (14.3)\\
\cline{1-7}
1&-0.25&1&10.78 (0.01) &10.77 (27.1) &10.78 (27.1)&10.82 (27.2) \\
\cline{3-7}
&&3&18.64 (0.01) &18.63 (27.2)&18.57 (27.1)&18.91 (27.6)\\
\cline{2-7}
&-0.125&1&7.28 (0.01)&7.28 (18.3)&7.28 (18.3)&7.31 (18.3)\\
\cline{3-7}
&& 3&12.65 (0.01)&12.64 (18.4)& 12.58 (18.3)& 12.68 (18.4)\\
\cline{2-7}
& -0.0625 & 1&6.02 (0.01)&6.02 (15.1)&6.01 (15.1)&6.03 (15.1)\\
\cline{3-7}
&& 3&10.45 (0.01) &10.43 (15.1)& 10.37 (15.0)& 10.47 (15.2) \\ \hline
\multicolumn{3}{|c|}{Average Rel Err \%} & 
& 0.1 & 0.4 & 0.4\\ \hline
\end{tabular}
\caption[The comparison of basket call option prices with MC, PEA, AEA and LBA methods. $\sigma _i(t,S)=0.2$.]{\label{table3.4} 
The comparison of  European basket call option prices  
with the Monte Carlo (MC), the
partial exact approximation (PEA), the asymptotic expansion approximation (AEA)
and the lower bound approximation (LBA).  The table displays the results with different jump intensities
$\lambda$, jump sizes $h_i=e^\eta-1$,  maturities $T$, and local volatility function 
 $\sigma_i(t,S)=0.2S$. 
The numbers inside brackets in the MC columns are the standard deviations
and those in the PEA, AEA and LBA columns are the implied volatilities (IV) computed with the Black-Scholes call option formula.}
}
\end{center}
\end{table}

\begin{table}
\begin{center}
{\small
\renewcommand{\arraystretch}{1.2}
\begin{tabular}{|c|c|rr|rr|}
\hline
\multicolumn{2}{|c|}{}&\multicolumn{2}{|c|}{Price}&\multicolumn{2}{|c|}{Imp Vol \%}\\
\hline
$T$ & Moneyness \% & MC (stdev) & LBA & MC & LBA\\ \hline
0.5& 70 & 32.31 (0.01) & 32.83 & 49.4 & 53.0\\
& 90 & 19.06 (0.02) & 19.60 & 50.6 & 52.7\\
& 100 & 14.26 (0.03) & 14.68 & 50.8 & 52.3\\
& 110 & 10.57 (0.01) & 10.8 & 51.1 & 51.9\\
& 130 & 5.63 (0.01) & 5.57 & 51.3 & 51.1 \\
\hline
2 & 70 & 37.11 (0.07) & 36.57 & 39.1 & 37.7\\
& 90 & 29.88 (0.10) & 28.99 & 46.7 & 44.9\\
& 100 & 27.02 (0.07) & 25.76 & 48.9 & 46.5\\
& 110 & 24.53 (0.08) & 22.85 & 50.5 & 47.4\\
& 130 & 20.44 (0.09) & 17.87 & 52.8 & 48.3\\
\hline
\multicolumn{2}{|c|}{Average Rel Err \%}&
&3.9& & 4.4\\
\hline
\end{tabular}
\caption[The comparison of basket call option prices with MC, PEA, AE and LB methods. $\sigma _i(t,S)=0.2$.]{\label{table2} 
The comparison of  European basket call option prices  
with the Monte Carlo (MC)
and the lower bound approximation (LBA).  The table displays the prices and their implied volatilities with different maturities $T$ and moneyness $K/S$.  The local volatility function 
is $\sigma_i(t,S)=0.5S$, the correlation coefficient of Brownian motions is $\rho=0.9$, the jump intensity $\lambda=4$, and jumpsizes $h_1=0, h_2=0.1, h_3=0.3, h_4=-0.5$.  
The numbers inside brackets in the MC columns are the standard deviations. The last row is the average relative errors between results of MC and LBA.}
}
\end{center}
\end{table}

\begin{table}
\begin{center}
{\small
\renewcommand{\arraystretch}{1.2}
\begin{tabular}{|c|r|r|rrr|}
\hline
$T$& $\alpha$ & $\beta$& MC (stdev) &AEA & LBA \\ \hline
1 &0.2& 1&7.35 (0.01)&7.35 &7.37 \\
&0.5&&14.71 (0.01)&14.42 &14.87 \\ \cline{2-6}
&0.2&0,8&5.31 (0.01)&5.33 & 5.31  \\
&0.5&&7.33 (0.01)&7.33 &7.34 \\ \cline{2-6}
&0.2&0.5&5.09 (0.01)&5.09 &5.08 \\
&0.5&&5.11 (0.01)&5.12 &5.11 \\ \hline
3&0.2&1&12.93 (0.01)&12.85 &12.86 \\
&0.5&&25.69 (0.04)&24.14 &26.16 \\ \cline{2-6}
&0.2&0.8&9.61 (0.01)&9.64 &9.63 \\
&0.5&&12.86 (0.01)&12.86 &12.81\\ \cline{2-6}
&0.2&0.5&8.96 (0.01)&8.98 &8.91\\
&0.5&&9.18 (0.01)&9.21 &9.18 \\ \hline
\multicolumn{3}{|c|}{Average Rel Err \%} & 
& 0.7& 0.4\\ \hline
\end{tabular}
\caption[ The comparison of  European basket call option prices with 
 the MC,  AEA, and LBA methods. $\lambda=0.3$]{\label{table3.1} 
 The comparison of  European basket call option prices with 
 the MC,  AEA and LBA.
The table displays results 
with different maturities $T$, local volatility functions 
$\sigma_i(t,S)=\alpha S^{\beta}$ and
jump sizes $h_i=e^{-0.25}-1 =-0.2212$. }
}
\end{center}
\end{table}

\begin{table}
\begin{center}
{\small
\renewcommand{\arraystretch}{1.2}
\begin{tabular}{|c|r|r|rr|}
\hline
$T$& $\alpha$ & $\beta$& MC (stdev) & LBA \\ \hline
1 &0.2& 1&5.53 (0.01) & 5.52 \\
&0.5&&13.87 (0.01)& 13.95 \\ \cline{2-5}
&0.2&0.8&2.22 (0.01)& 2.22  \\
&0.5&&5.50 (0.01)&5.49 \\ \cline{2-5}
&0.2&0.5&0.63 (0.01)&0.63 \\
&0.5&&1.42 (0.01)& 1.42 \\ \hline
3&0.2&1&9.68 (0.02)&9.66  \\
&0.5&&24.42 (0.06)&24.84 \\ \cline{2-5}
&0.2&0.8&3.95 (0.01)&3.94 \\
&0.5&&9.57 (0.02)& 9.59 \\ \cline{2-5}
&0.2&0.5&1.37 (0.01)&1.37 \\
&0.5&&2.59 (0.01)& 2.59 \\ \hline
\multicolumn{3}{|c|}{Average Rel Err \%} & 
 & 0.3\\ \hline
\end{tabular}
\caption[ The comparison of  European basket call option prices with 
 the MC and LB methods. $\lambda=0.3$]{\label{table3.3} 
 The comparison of  European basket call option prices with 
 the MC and LBA.
The table displays results 
with different maturities $T$, local volatility functions 
$\sigma_i(t,S)=\alpha S^{\beta}$, and 
jump sizes $h_1=0$, $h_2=0.3$, $h_3=-0.3$, and $h_4=0$. }
}
\end{center}
\end{table}

\end{document}